# Some new solutions to the Klein-Gordon equation in 1-1D space-time in the presence of time dependent potentials.

Dan Solomon
Rauland-Borg Corporation

Email: dan.solomon@rauland.com
July 12, 2010

**Abstract.**

We find three exact solutions to the Klein-Gordon equation in 1-1 dimension space-time for different time dependent potentials. In two cases we consider a time dependent scalar potential and in one case a time dependent electric potential.

## 1. Introduction.

The Klein-Gordon equation in 1-1D space-time in the presence of a scalar potential $V_s(x,t)$ and electric potential $V_e(x,t)$ is given by,

$$\left(\frac{\partial}{\partial t} - iV_e\right)\left(\frac{\partial}{\partial t} - iV_e\right)\varphi - \frac{\partial^2 \varphi}{\partial x^2} + \left(m^2 + V_s\right)\varphi = 0 \qquad (1.1)$$

where $m$ is the mass. There are a number of solutions that appear in the literature for various types and combinations of scalar and electric potentials [1-9]. In almost all cases these potentials are time independent.

In this paper we will find solutions to this equation for three different potentials that are dependent on both space and time. In Section 2 we will solve (1.1) for the potential given by,

$$V_e(x,t) = 0 \text{ and } V_s(x,t) = \chi_1'(x+t) - \chi_2'(x-t) - \frac{\chi_1'(x+t)\chi_2'(x-t)}{m^2} \qquad (1.2)$$

where $\chi_1$ and $\chi_2$ are arbitrary functions and,

$$\chi_i'(u) \equiv \frac{d\chi_i(u)}{du} \qquad (1.3)$$

In Section 3 we solve (1.1) for the case where $m = 0$ and the potential is a time dependent delta function given by,

$$V_e(x,t) = 0 \text{ and } V_s(x,t) = 2\lambda(t)\delta(x) \tag{1.4}$$

In Section 4 we will consider the case where $m = 0$ and,

$$V_e(x,t) = 2\lambda(t)\theta(x); \quad V_s(x,t) = 0 \tag{1.5}$$

where $\theta(x)$ is the step function which is defined by,

$$\theta(x) = \begin{cases} 0 \text{ for } x < 0 \\ 1 \text{ for } x \geq 0 \end{cases} \tag{1.6}$$

## 2. A solution for a time-dependent scalar potential.

If $V_e(x,t) = 0$ the Klein-Gordon equation is given by,

$$\frac{\partial^2 \varphi}{\partial t^2} - \frac{\partial^2 \varphi}{\partial x^2} + (m^2 + V_s)\varphi = 0 \tag{2.1}$$

For $V_s(x,t)$ given by (1.2) the solution to the above equation is,

$$\varphi(x,t) = \varphi_p^{(0)}(x,t) \exp\left[i\left(\frac{\chi_1(x+t)}{2(E+p)} + \frac{\chi_2(x-t)}{2(E-p)}\right)\right] \tag{2.2}$$

where,

$$\varphi_p^{(0)}(x,t) = e^{iEt}e^{-ipx} \tag{2.3}$$

with,

$$E = \pm\sqrt{p^2 + m^2} \tag{2.4}$$

This will be proven as follows. Use (2.2) to obtain,

$$\frac{\partial^2 \varphi}{\partial t^2} = \left\{i\left(\frac{\chi_1''(x+t)}{2(E+p)} + \frac{\chi_2''(x-t)}{2(E-p)}\right) - \left(\frac{\chi_1'(x+t)}{2(E+p)} - \frac{\chi_2'(x-t)}{2(E-p)} + E\right)^2\right\}\varphi \tag{2.5}$$

$$\frac{\partial^2 \varphi}{\partial x^2} = \left\{i\left(\frac{\chi_1''(x+t)}{2(E+p)} + \frac{\chi_2''(x-t)}{2(E-p)}\right) - \left(\frac{\chi_1'(x+t)}{2(E+p)} + \frac{\chi_2'(x-t)}{2(E-p)} - p\right)^2\right\}\varphi \tag{2.6}$$

From this we obtain,

$$\frac{\partial^2 \varphi}{\partial t^2} - \frac{\partial^2 \varphi}{\partial x^2} = -\left\{\left(\frac{\chi_1'(x+t)}{2(E+p)} - \frac{\chi_2'(x-t)}{2(E-p)} + E\right)^2 - \left(\frac{\chi_1'(x+t)}{2(E+p)} + \frac{\chi_2'(x-t)}{2(E-p)} - p\right)^2\right\}\varphi \tag{2.7}$$

This yields,

$$\frac{\partial^2 \varphi}{\partial t^2} - \frac{\partial^2 \varphi}{\partial x^2} + m^2 \varphi = \left\{ \frac{\chi_1'(x+t)\chi_2'(x-t)}{m^2} - \chi_1'(x+t) + \chi_2'(x-t) \right\} \varphi \qquad (2.8)$$

where we have used $(E-p)(E+p) = m^2$ to obtain the above result. Use (1.2) in this result to obtain (2.1). This completes the proof.

### 3. A time dependent delta function potential.

Consider the massless case, i.e., $m = 0$, and let $V_e(x,t) = 0$ and the scalar potential be given by (1.4). In this case the Klein-Gordon equation becomes,

$$\frac{\partial^2 \varphi}{\partial t^2} - \frac{\partial^2 \varphi}{\partial x^2} = -2\lambda(t)\delta(x)\varphi \qquad (3.1)$$

The equation can be solved by making note of the fact that for $x \ne 0$,

$$\frac{\partial^2 \varphi}{\partial t^2} - \frac{\partial^2 \varphi}{\partial x^2}\bigg|_{x \ne 0} = 0 \qquad (3.2)$$

The situation at $x = 0$ is handled as follows. Let $\varepsilon > 0$ and let $\varepsilon \to 0$. Assume $\varphi(x,t)$ is continuous at $x = 0$ so that,

$$\varphi(+\varepsilon, t) \underset{\varepsilon \to 0}{=} \varphi(-\varepsilon, t) \qquad (3.3)$$

Integrate (3.1) from $x = -\varepsilon$ to $x = +\varepsilon$ to obtain,

$$\frac{\partial^2}{\partial t^2}\int_{-\varepsilon}^{+\varepsilon}\varphi(x,t)dx - \left(\frac{\partial\varphi(x,t)}{\partial x}\bigg|_{x=+\varepsilon} - \frac{\partial\varphi(x,t)}{\partial x}\bigg|_{x=-\varepsilon}\right) = -2\lambda(t)\varphi(0,t) \qquad (3.4)$$

In the limit $\varepsilon \to 0$ we have $\int_{-\varepsilon}^{+\varepsilon}\varphi(x,t)dx \underset{\varepsilon \to 0}{=} \varphi(0,t)\varepsilon$. Use this in (3.4) to obtain,

$$\varepsilon\frac{\partial^2\varphi(0,t)}{\partial t^2} - \left(\frac{\partial\varphi(x,t)}{\partial x}\bigg|_{x=+\varepsilon} - \frac{\partial\varphi(x,t)}{\partial x}\bigg|_{x=-\varepsilon}\right) \underset{\varepsilon \to 0}{=} -2\lambda(t)\varphi(0,t) \qquad (3.5)$$

In the limit $\varepsilon \to 0$ the quantity $\varepsilon\,\partial^2\varphi(0,t)/\partial t^2 \to 0$ as long as $\partial^2\varphi(0,t)/\partial t^2$ is non-singular. Assuming this is the case (3.5) becomes,

$$\frac{\partial\varphi(x,t)}{\partial x}\bigg|_{x=+\varepsilon} - \frac{\partial\varphi(x,t)}{\partial x}\bigg|_{x=-\varepsilon} \underset{\varepsilon \to 0}{=} 2\lambda(t)\varphi(0,t) \qquad (3.6)$$

We consider two forms of the solution,

$$\varphi_1(x,t) = \begin{cases} e^{ik(t-x)} + B_1(t+x)e^{ik(t+x)} & \text{for } x < 0 \\ A_1(t-x)e^{ik(t-x)} & \text{for } x > 0 \end{cases} \qquad (3.7)$$

and,

$$\varphi_2(x,t) = \begin{cases} A_2(t+x)e^{ik(t+x)} & \text{for } x < 0 \\ e^{ik(t+x)} + B_2(t-x)e^{ik(t-x)} & \text{for } x > 0 \end{cases} \qquad (3.8)$$

The first solution (3.7) can be viewed as a wave $e^{ik(t-x)}$ moving in the positive x-direction and being scattered by the scalar potential at $x = 0$. The quantity $B_1(t+x)e^{ik(t+x)}$ is the reflected wave and the $A_1(t-x)e^{ik(t-x)}$ is the transmitted wave. Similarly (3.8) represents the scattering of a wave moving in the negative x-direction.

We will first solve for $\varphi_1$. As can be readily shown $\varphi_1(x,t)$ satisfies Eq. (3.2). Next use (3.3) to obtain,

$$1 + B_1(t) = A_1(t) \qquad (3.9)$$

and use (3.6) to yield,

$$\left(-\frac{dA_1(t)}{dt} - ikA_1(t)\right) - \left(-ik + \frac{dB_1(t)}{dt} + ikB_1(t)\right) = 2\lambda(t)A_1(t) \qquad (3.10)$$

Note in deriving the above we have used $df(t \pm x)/dx \underset{x=0}{=} \pm df(t)/dt$. Use (3.9) in (3.10) to obtain,

$$\frac{dB_1(t)}{dt} + B_1(t)(ik + \lambda(t)) = -\lambda(t) \qquad (3.11)$$

The condition that $\partial^2\varphi(0,t)/\partial t^2$ is non-singular is true if $\partial\varphi(0,t)/\partial t$ is continuous. From the above equations it is evident that this is true if $\lambda(t)$ is continuous. In the following discussion we will assume that this is the case.

Eq. (3.11) can be rewritten as,

$$\frac{d}{dt}\left(B_1(t)e^{(ikt+F(t,t_0,[\lambda]))}\right) = -\lambda(t)e^{(ikt+F(t,t_0,[\lambda]))} \qquad (3.12)$$

where,

$$F(t,t_0,[\lambda]) = \int_{t_0}^{t} \lambda(t')dt' \qquad (3.13)$$

where $t_0$ is an arbitrary constant. Integrate this to obtain,

$$\left(B_1(t)e^{(ikt+F(t,t_0,[\lambda]))}\right) - \left(B_1(t_1)e^{(ikt_1+F(t_1,t_0,[\lambda]))}\right) = -\int_{t_1}^{t} \lambda(t)e^{(ikt+F(t,t_0,[\lambda]))} \qquad (3.14)$$

where $B_1(t_1)$ is the initial condition on $B_1$ at time $t = t_1$. Rearrange terms to obtain,

$$B_1(t) = e^{-(ikt+F(t,t_0,[\lambda]))}\left\{ B_1(t_1)e^{(ikt_1+F(t_1,t_0,[\lambda]))} - \int_{t_1}^{t} dt'\left[\lambda(t')e^{(ikt'+F(t',t_0,[\lambda]))}\right]\right\} \qquad (3.15)$$

Therefore we can solve for $B_1(t)$ if we know the initial condition $B_1(t_1)$ and can do the integrals. From this result we can readily obtain $B_1(t+x)$ and, using (3.9), obtain $A_1(t-x)$. Apply these results to (3.7) to solve for $\varphi_1(x,t)$.

When we solve for $\varphi_2(x,t)$ (Eq. (3.8)) we find that the equations for $A_2$ and $B_2$ are identical to that of $A_1$ and $B_1$. Recall that $B_1$ and $B_2$ are the coefficients of the reflected wave. If we consider problems where the initial conditions on $B_1$ and $B_2$ are identical we have,

$$B_2(t) = B_1(t) \qquad (3.16)$$

It follows from this that,

$$A_2(t) = A_1(t) \qquad (3.17)$$

Using this result along with (3.7) and (3.8) we can define even solutions by the expression $\psi_+(x,t) = (\varphi_1(x,t) + \varphi_2(x,t))/2$ and odd solutions by $\psi_-(x,t) = (\varphi_2(x,t) - \varphi_1(x,t))/2i$. Using these definitions and (3.17) and (3.16) along with (3.7) and (3.8) we obtain,

$$\psi_+(x,t) = e^{ikt}\begin{cases} \cos(kx) + B_1(t+x)e^{ikx} & \text{for } x < 0 \\ \cos(kx) + B_1(t-x)e^{-ikx} & \text{for } x > 0 \end{cases} \qquad (3.18)$$

and,

$$\psi_-(x,t) = e^{ikt}\sin(kx) \qquad (3.19)$$

Note that the odd solution $\psi_-(x,t)$ is not affected by the scalar potential $V_s(x,t) = 2\lambda(t)\delta(x)$. This is because $\psi_-(x,t)$ is zero when $x = 0$.

For an example of a solution let,

$$\lambda(t) = \left(\frac{2t}{(t^2+a)}\right)\theta(t) \tag{3.20}$$

Let $B_1(0) = 0$ for $t < 0$. To solve (3.15) the initial condition is $B_1(0) = 0$ and the initial time is $t_1 = 0$. We also set $t_0 = 0$. This yields $F(t, 0, [\lambda]) = \log\left[(t^2+a)/a\right]$. Use all this in (3.15) to obtain,

$$B_1(t)\Big|_{t\geq 0} = -\left(\frac{2e^{-ikt}}{(t^2+a)}\right)\left\{\int_0^t t' e^{ikt'} dt'\right\} \tag{3.21}$$

This can be integrated out to yield,

$$B_1(t) = \theta(t)\left(\frac{-2}{(t^2+a)}\right)\left\{\frac{(1-e^{-ikt})}{k^2} - \frac{it}{k}\right\} \tag{3.22}$$

## 4. A time dependent step electric potential.

In this section we will solve for the zero mass Klein-Gordon equation in the presence of a time dependent step electric potential as specified by (1.5). The zero mass Klein-Gordon equation in the presence of an electric potential $V_e(x,t)$ is,

$$\left(\frac{\partial}{\partial t} - iV_e\right)\left(\frac{\partial}{\partial t} - iV_e\right)\varphi - \frac{\partial^2 \varphi}{\partial x^2} = 0 \tag{4.1}$$

This can be rewritten as,

$$\frac{\partial^2 \varphi}{\partial t^2} - 2iV_e \frac{\partial \varphi}{\partial t} - i\frac{\partial V_e}{\partial t}\varphi - V_e^2\varphi - \frac{\partial^2 \varphi}{\partial x^2} = 0 \tag{4.2}$$

Let $V_e(x,t) = 2\theta(x)\lambda(t)$. The solution to (4.1) is then,

$$\varphi_1(x,t) = \left\langle \begin{array}{l} e^{ik(t-x)} + D_1(t+x)e^{ik(t+x)} \text{ for } x < 0 \\ C_1(t-x)e^{ik(t-x)} \exp\left(2i\int_{-\infty}^{t} \lambda(t')dt'\right) \text{ for } x > 0 \end{array} \right. \tag{4.3}$$

and,

$$\varphi_2(x,t) = \begin{cases} C_2(t+x)e^{ik(t+x)} & \text{for } x < 0 \\ \left(e^{ik(t+x)} + D_2(t-x)e^{ik(t-x)}\right)\exp\left(2i\int_{-\infty}^{t}\lambda(t')dt\right) & \text{for } x > 0 \end{cases} \qquad (4.4)$$

At $x=0$, $\varphi(x,t)$ and its first derivative with respect to $x$ is continuous. This yields,

$$1 + D_1(t) = C_1(t)\exp\left(2i\int_{-\infty}^{t}\lambda(t')dt\right) \qquad (4.5)$$

and,

$$\frac{\partial D_1(t)}{\partial t} + ik(D_1(t)-1) = -\left\{\frac{\partial C_1(t)}{\partial t} + ikC_1(t)\right\}\exp\left(2i\int_{-\infty}^{t}\lambda(t')dt\right) \qquad (4.6)$$

Use (4.5) in (4.6) to obtain,

$$\frac{\partial D_1(t)}{\partial t} + iD_1(t)(k - \lambda(t)) = i\lambda(t) \qquad (4.7)$$

This equation is the same as (3.11) with $\lambda(t)$ replaced by $-i\lambda(t)$. Therefore the solution to (4.7) is,

$$D_1(t) = e^{-(ikt + F(t,t_0,[-i\lambda]))}\left\{D_1(t_1)e^{(ikt_1 + F(t_1,t_0,[-i\lambda]))} + i\int_{t_1}^{t}dt'\left[\lambda(t')e^{(ikt' + F(t',t_0,[-i\lambda]))}\right]\right\} \qquad (4.8)$$

Next we will solve for $\varphi_2(x,t)$. In this case the boundary conditions yields,

$$C_2(t) = (1 + D_2(t))\exp\left(2i\int_{-\infty}^{t}\lambda(t')dt\right) \qquad (4.9)$$

and

$$\frac{\partial C_2(t)}{\partial t} + ikC_2(t) = \left(ik - \frac{\partial D_2(t)}{\partial t} - ikD_2(t)\right)\exp\left(2i\int_{-\infty}^{t}\lambda(t')dt\right) \qquad (4.10)$$

Use (4.9) in the above to obtain,

$$\frac{\partial D_2(t)}{\partial t} + iD_2(t)(k + \lambda(t)) = -i\lambda(t) \qquad (4.11)$$

Note that this is equivalent to (4.7) with $\lambda(t)$ replaced by $-\lambda(t)$. Therefore solution to (4.11). is given by (4.8) with $D_1$ replaced by $D_2$ and $\lambda(t)$ replaced by $-\lambda(t)$.

In conclusion we have found the solutions to the Klein-Gordon equation for three different time dependent potentials.